\documentclass[preprint,12pt]{elsarticle}

\usepackage{amssymb}
\usepackage{lipsum}
\usepackage{amsmath}
\usepackage{hyperref} 
\usepackage[mathlines]{lineno} 
\usepackage[dvipsnames]{xcolor}
\usepackage{graphicx} 

\def\gtorder{\mathrel{\raise.3ex\hbox{$>$}\mkern-14mu
 \lower0.6ex\hbox{$\sim$}}}
\def\ltorder{\mathrel{\raise.3ex\hbox{$<$}\mkern-14mu
 \lower0.6ex\hbox{$\sim$}}}

\def\gtorder{\mathrel{\raise.3ex\hbox{$>$}\mkern-14mu
 \lower0.6ex\hbox{$\sim$}}}
\def\ltorder{\mathrel{\raise.3ex\hbox{$<$}\mkern-14mu
 \lower0.6ex\hbox{$\sim$}}}

\usepackage{etoolbox}
\journal{Physics Letters B}

\begin{document}

\begin{frontmatter}


\title{Inclusive studies of two- and three-nucleon short-range correlations in $^3$H and $^3$He}

\author[unh,lbl]{S.~Li\corref{cor1} }  

\address[unh]{University of New Hampshire, Durham, New Hampshire 03824, USA}
\address[lbl]{Lawrence Berkeley National Laboratory, Berkeley, California 94720, USA}
\address[anl]{Physics Division, Argonne National Laboratory, Lemont, Illinois 60439, USA} 
\address[mit]{Massachusetts Institute of Technology, Cambridge, Massachusetts 02139, USA} 
 \address[uva]{University of Virginia, Charlottesville, Virginia 22904, USA}
 \address[kent]{Kent State University, Kent, Ohio 44240, USA}
 \address[king_saud]{King Saud University, Riyadh 11451, Kingdom of Saudi Arabia}
\address[Zagreb]{University of Zagreb, Zagreb, Croatia} 

\author[unh]{S.~N.~Santiesteban}

\author[lbl,anl]{J.~Arrington} 

\author[mit]{R.~Cruz-Torres} 

\author[unh]{L.~Kurbany}

\author[uva]{D.~Abrams}

\author[kent,king_saud]{S.~Alsalmi }

\author[Zagreb]{D.~Androic} 
\author[cal_la]{K.~Aniol} \address[cal_la]{California State University, Los Angeles, California 90032, USA} 

\author[wm]{T.~Averett} \address[wm]{The College of William and Mary, Williamsburg, Virginia 23185, USA} 

\author[wm]{C.~Ayerbe~Gayoso} 

\author[utk]{J.~Bane} \address[utk]{University of Tennessee, Knoxville, Tennessee 37966, USA} 

\author[wm]{S.~Barcus}

\author[utk]{J.~Barrow} 

\author[mit]{A.~Beck} 

\author[infn_catania]{V.~Bellini} \address[infn_catania]{INFN Sezione di Catania, Italy} 

\author[missi]{H.~Bhatt} \address[missi]{Mississippi State University, Mississippi State, Mississippi 39762, USA} 

\author[missi]{D.~Bhetuwal}

\author[hampton]{D.~Biswas} \address[hampton]{Hampton University, Hampton, Virginia 23669, USA} 

\author[odu]{D.~Bulumulla} \address[odu]{Old Dominion University, Norfolk, Virginia 23529, USA} 

\author[jlab]{A.~Camsonne} \address[jlab]{Thomas Jefferson National Accelerator Facility, Newport News, Virginia 23606, USA} 

\author[fiu]{J.~Castellanos} \address[fiu]{Florida International University, Miami, Florida 33199, USA} 

\author[wm]{J.~Chen}

\author[jlab]{J-P.~Chen} 

\author[msu]{D.~Chrisman} \address[msu]{Michigan State University, East Lansing, Michigan 48824, USA} 

\author[hampton,jlab]{M.~E.~Christy} 

\author[sbu]{C.~Clarke} \address[sbu]{Stony Brook, State University of New York, New York 11794, USA} 

\author[jlab]{S.~Covrig} 

\author[utk]{K.~Craycraft} 

\author[uva]{D.~Day} 

\author[missi]{D.~Dutta} 

\author[uconn]{E.~Fuchey} \address[uconn]{University of Connecticut, Storrs, Connecticut 06269, USA} 

\author[uva]{C.~Gal} 

\author[infn_rome]{F.~Garibaldi} \address[infn_rome]{INFN, Rome, Italy} 

\author[hampton]{T.~N.~Gautam} 

\author[tohoku]{T.~Gogami} \address[tohoku]{Tohoku University, Sendai, Japan} 

\author[jlab]{J.~Gomez} 

\author[hampton,msu]{P.~Gu\`eye} 

\author[hampton]{A.~Habarakada} 

\author[kent]{T.~J.~Hague} 

\author[jlab]{J.~O.~Hansen} 

\author[odu]{F.~Hauenstein}  

\author[temple]{W.~Henry} \address[temple]{Temple University, Philadelphia, Pennsylvania 19122, USA} 

\author[jlab]{D.~W.~Higinbotham} 

\author[caltech,anl]{R.~J.~Holt} \address[caltech]{California Institute of Technology, Pasadena, California 91125, USA}

\author[odu]{C.~E.~Hyde} 

\author[tohoku]{K.~Itabashi}

\author[tohoku]{M.~Kaneta}

\author[missi]{A.~Karki} 

\author[kent]{A.~T.~Katramatou} 

\author[jlab]{C.~E.~Keppel}

\author[odu]{M.~Khachatryan} 

\author[sbu]{V.~Khachatryan} 

\author[ohio]{P.~M.~King} \address[ohio]{Ohio University, Athens, Ohio 45701, USA} 

\author[negev]{I.~Korover} \address[negev]{Nuclear Research Center-Negev, Beer-Sheva, Israel} 

\author[sbu]{T.~Kutz}

\author[hampton]{N.~Lashley-Colthirst} 

\author[wm]{W.~B.~Li} 

\author[columbia]{H.~Liu} \address[columbia]{Columbia University, New York, New York 10027, USA} 

\author[uva]{N.~Liyanage} 

\author[unh]{E.~Long} 

\author[manitoba]{J.~Mammei} \address[manitoba]{University of Manitoba, Winnipeg, MB R3T 2N2, Canada} 

\author[fiu]{P.~Markowitz} 

\author[jlab]{R.~E.~McClellan} 

\author[infn_rome]{F.~Meddi} 

\author[jlab]{D.~Meekins} 

\author[mit]{S.~Mey-Tal Beck}

\author[jlab]{R.~Michaels} 

\author[jsi,lju,jgu]{M.~Mihovilovi\v{c}}
\address[jsi]{Jo\v{z}ef Stefan Institute, 1000 Ljubljana, Slovenia} 
\address[lju]{Faculty of Math and Physics, University of Ljubljana, 1000 Ljubljana, Slovenia} 
\address[jgu]{Institut f\"{u}r Kernphysik, Johannes Gutenberg-Universit\"{a}t Mainz, Mainz, Germany} 

\author[cnu]{A.~Moyer} \address[cnu]{Christopher Newport University, Newport News, Virginia 23606, USA} 

\author[tohoku]{S.~Nagao} 

\author[uva]{V.~Nelyubin}

\author[uva]{D.~Nguyen} 

\author[kent]{M.~Nycz}

\author[snc]{M.~Olson} \address[snc]{Saint Norbert College, De Pere, Wisconsin 54115, USA} 

\author[mit]{L.~Ou}

\author[wm]{V.~Owen} 

\author[uva]{C.~Palatchi} 

\author[hampton]{B.~Pandey}       \fnref{pandey_address} 

\author[mit]{A.~Papadopoulou} 

\author[sbu]{S.~Park}

\author[wm]{S.~Paul}   

\author[zagreb]{T.~Petkovic} 

\author[kharkov]{R.~Pomatsalyuk} \address[kharkov]{Institute of Physics and Technology, Kharkov, Ukraine} 

\author[uva]{S.~Premathilake} 

\author[norfolk]{V.~Punjabi} \address[norfolk]{Norfolk State University, Norfolk, Virginia 23529, USA} 

\author[rutgers]{R.~D.~Ransome} \address[rutgers]{Rutgers University, New Brunswick, New Jersey 08854, USA} 

\author[anl]{P.~E.~Reimer} 

\author[fiu]{J.~Reinhold} 

\author[anl]{S.~Riordan}

\author[ohio]{J.~Roche} 

\author[mendez]{V.~M. Rodriguez} \address[mendez]{Divisi\'{o}n de Ciencias y Tecnolog\'{i}a, Universidad Ana G. M\'{e}ndez, Recinto de Cupey, San Juan 00926, Puerto Rico} 

\author[mit]{A.~Schmidt}

\author[mit]{B.~Schmookler}

\author[mit]{E.~P.~Segarra}

\author[yerevan]{A.~Shahinyan} \address[yerevan]{Yerevan Physics Institute, Yerevan, Armenia}

\author[lju,jsi]{S.~\v{S}irca}

\author[unh]{K.~Slifer}

\author[unh]{P.~Solvignon}

\author[kent]{T.~Su} 

\author[jlab]{R.~Suleiman}

\author[jlab]{H.~Szumila-Vance}

\author[jlab]{L.~Tang} 

\author[syracuse]{Y.~Tian} \address[syracuse]{Syracuse University, Syracuse, New York 13244, USA} 

\author[nmu]{W.~Tireman} \address[nmu]{Northern Michigan University, Marquette, Michigan 49855, USA}

\author[infn_c]{F.~Tortorici}

\author[tohoku]{Y.~Toyama} 

\author[tohuku]{K.~Uehara} 

\author[infn_rome]{G.~M.~Urciuoli} 

\author[msu]{D.~Votaw} 

\author[glasgow]{J.~Williamson} \address[glasgow]{University of Glasgow, Glasgow, G12 8QQ Scotland, UK} 

\author[jlab]{B.~Wojtsekhowski} 

\author[jlab]{S.~Wood}
 
\author[thu,anl]{Z.~H.~Ye}
\address[thu]{Tsinghua University, Beijing, China}

\author[uva]{J.~Zhang} 

\author[uva]{X.~Zheng} 

\cortext[cor1]{Corresponding author: ShujieLi@lbl.gov}

\fntext[pandey_address]{Current address: Department of Physics \& Astronomy, Virginia Military Institute, Lexington, Virginia 24450, USA}
\fntext[Paul_address]{Current address: University of California Riverside, 900 University Ave. Riverside, CA 92521, USA}

\begin{abstract}

Inclusive electron scattering at carefully chosen kinematics can isolate scattering from the high-momentum nucleons in short-range correlations (SRCs). SRCs are produced by the hard, short-distance interactions of nucleons in the nucleus, and because the two-nucleon (2N) SRCs arise from the same N-N interaction in all nuclei, the cross section in the SRC-dominated regime is identical up to an overall scaling factor. This scaling behavior has been used to identify SRC dominance and to measure the contribution of SRCs in a wide range of nuclei. We examine this scaling behavior over a range of momentum transfers using new data on $^2$H, $^3$H, and $^3$He, and find an expanded scaling region compared to heavy nuclei. Motivated by this improved scaling, we examine the $^3$H and $^3$He data in kinematics where three-nucleon SRCs may play an important role. The data for the largest struck nucleon momenta are consistent with isolation of scattering from three-nucleon SRCs, and suggest that the very highest momentum nucleons in $^3$He have a nearly isospin-independent momentum configuration. 

\end{abstract}

\end{frontmatter}

In a shell-model or mean-field (MF) picture of a nucleus, typical nucleon momenta are near or below the Fermi momentum of the nucleus, $k_F \approx 250$~MeV/c. The strong, short-range part of the nucleon-nucleon (N-N) interaction generates two-nucleon short-range correlations (2N-SRCs), pairs of high momentum nucleons with a large relative momentum ($\gtorder 2k_F$), and smaller total momentum (consistent with the MF picture)~\cite{frankfurt88, Arrington:2022sov}. This mechanism brings up to $\sim$20\% of nucleons in a nucleus to momenta higher than the typical maximum MF momentum of the nucleus, $k_{MF}$~\cite{Arrington:2022sov}. Inclusive quasi-elastic (QE) scattering can isolate these high-momentum configurations. Measurements at large momentum transfer ($Q^2$) suppress long-range final-state interactions (FSIs), meson-exchange currents, and isobar configurations~\cite{frankfurt88, sargsian03, arrington12a-SRCreview}, while inelastic contributions are suppressed at low energy transfer ($\nu$), corresponding to $x = Q^2/(2M_p\nu) > 1$, where $M_p$ is the proton mass.

In QE scattering from a deuteron, any given combination of $x$ and $Q^2$ corresponds to a minimum initial momentum of the struck nucleon, $k_{min}$, below which scattering is kinematically forbidden~\cite{sargsian03}. For a nucleus A, a similar $k_{min}$ threshold can be defined for QE scattering, which differs slightly from the value for the deuteron due to the heavier spectator system. For all nuclei, the QE peak at $x \approx 1$ is dominated by scattering from low-momentum nucleons. At $x > 1$, As $x$ and $Q^2$ increase, $k_{min}$ also increases, leading to a rapid decrease in the scattering from MF nucleons, eventually leaving scattering from high-momentum SRC pairs as the dominant process for $k_{min}>k_{MF}$~\cite{frankfurt88, sargsian03, Arrington:2022sov}. If we assume that SRCs in nuclei are identical to those in the deuteron, i.e. unbound $np$ pairs with zero total momentum, then the QE cross sections for scattering from different nuclei in this kinematic region will be identical up to an overall scaling factor representing the relative abundance of SRC pairs in the nucleus. As such, the cross section ratio of any nucleus $A$ to the deuteron, $\sigma_A/\sigma_{^2H}$, should scale with respect to both $x$ and $Q^2$ in the 2N-SRC dominated region. Contributions from a stationary 2N-SRC are limited to $x<2$, as $x = M_{^2H}/M_p \approx 2$ is the kinematic limit for scattering from a deuteron.

An examination of the A/$^2$H cross section ratios for light and heavy nuclei from SLAC data~\cite{frankfurt93} observed a plateau at $x>1.4$ that was independent of $Q^2$ for $Q^2 \ge 1.4$~GeV$^2$. The height of the plateau, $a_2(A)$, in this scaling regime has been measured for a variety of nuclei in this and later experiments~\cite{fomin12, schmookler19, Li2022}. These data showed that $a_2$ increases rapidly with A in light nuclei, but becomes roughly constant, $a_2 \approx 5$, in nuclei from carbon to lead~\cite{Arrington:2022sov}. As $Q^2$ decreases, larger $x$ values are required to obtain the initial nucleon momenta necessary to isolate SRCs, with scaling breaking down completely at lower $Q^2$ values~\cite{frankfurt93, egiyan03, fomin17} due to increased FSI contributions~\cite{arrington12a-SRCreview, fomin17, Arrington:2022sov}, discussed below. 

While the SRC picture described above predicted the observed scaling behavior of the A/$^2$H ratios, it makes several important assumptions: It neglects FSIs and treats all SRCs as deuteron-like and stationary~\cite{frankfurt93, sargsian03, Arrington:2022sov}. Calculations~\cite{Atti1994, Benhar1995} show that FSIs decrease rapidly with increasing $Q^2$ at low $Q^2$ values. In addition, above $Q^2 \gtorder 1$~GeV$^2$, FSIs are dominated by rescattering within the SRC pairs~\cite{Atti1994, Frankfurt2008, arrington12a-SRCreview} and should thus cancel in the A/$^2$H ratios in the SRC-dominated regime~\cite{sargsian03, arrington12a-SRCreview}. The dominance of deuteron-like SRCs has been shown in a series of measurements~\cite{Shneor:2007tu, Subedi:2008zz, Korover:2014dma, CLAS:2018xvc, Nguyen:2020mgo, Li2022}, which found that $pp$- and $nn$-SRCs contribute only a few percent each in heavy nuclei~\cite{Arrington:2022sov}, while $^3$He and $^3$H have a larger $pp$ or $nn$ contribution but are still dominated by np pairs at the 80+\% level~\cite{Li2022}. This observed $np$ dominance is understood to be a consequence of the nucleon-nucleon tensor force~\cite{schiavilla07, alvioli08, wiringa14, Terashima:2018bwq} that dominates the generation of high-momentum nucleons for $k > k_{MF}$.

CM motion and binding of the SRC pairs in the nucleus can also modify the simple picture of universal SRCs in all nuclei. Estimates show that the CM motion enhances the SRC contribution in the high momentum tails, enhancing the ratio in the plateau region, increasing $a_2$(A) by $\sim$10-20\% above the expectation based on simply taking the relative number of SRCs~\cite{fomin12, arrington12b-EMCSRC, Weiss:2020bkp}. While calculations show some $x$ dependence in this enhancement, in particular for $x \to 2$, the observation of a plateau over a range of $x$ and $Q^2$ values for several nuclei suggests that it does not significantly distort the $x$ and $Q^2$ independence in the plateau region~\cite{fomin12}. Binding is estimated to reduce $a_2$(A) but have a smaller effect~\cite{Weiss:2020bkp}. 

The need to isolate SRCs from the MF nucleons, coupled with the need to minimize FSIs and other scaling-violating effects, defines the region where the predicted scaling is observed in A/$^2$H ratio. This scaling region may be extended when examining ratios of similar nuclei, e.g. for $^3$He/$^2$H, $^3$H/$^3$He~\cite{Li2022} or $^{48}$Ca/$^{40}$Ca~\cite{Nguyen:2020mgo}. In these cases, FSI corrections and CM motion of the SRCs in the two nuclei are very similar, and so their effects on the cross sections will have significant cancellation in the target ratios. Comparing light nuclei has additional benefits from their smaller sizes of FSIs and CM motion, and faster falloff of the MF contributions due to their smaller MF momenta. All of those allow 2N-SRCs to dominate at lower values of $x$ and $Q^2$. 

At $x>2$, where 2N-SRCs only contribute due to their CM motion, the SRC picture predicts that scattering from three-nucleon SRCs (3N-SRCs) may begin to dominate~\cite{frankfurt93, Ye:2018jth, Arrington:2022sov, Fomin:2023gdz}. The 3N-SRC can be taken as a strongly-interacting, short-distance configuration where all three nucleons have momenta above $k_{MF}$ pending the CM motion. The inclusive measurement cannot directly distinguish between scattering from high momentum nucleons in 2N or 3N-SRC configurations. But if 3N-SRCs begin to contribute beyond $x=2$, the inclusive cross section ratio will change from the 2N-SRC region, yielding a second plateau at $x>2$ where the ratio reflects the relative contribution of 3N-SRCs~\cite{frankfurt93}.

The conventional approach to searching for 3N-SRC scaling is to measure the A/$^3$He cross section ratios at $x>2$ and large $Q^2$, to allow for the observation of a clear plateau at two or more $Q^2$ values~\cite{Fomin:2023gdz}, confirming both $x$ and $Q^2$ independence. Because inclusive scattering cannot separate scattering from 2N- or 3N-SRCs, this approach relies on observing a change in the A/$^3$He ratio at $x>2$ associated with the transition to 3N-SRC dominance~\cite{frankfurt93, Day:2018nja}.
Existing A/$^3$He data at $Q^2 \approx $1.5--2~GeV$^2$~\cite{ye18} did not observe the predicted 3N-SRC plateau, while a measurement at $Q^2 \approx 2.8$~GeV$^2$~\cite{fomin12} is consistent with a plateau for $x>2.5$, but with very large uncertainties~\cite{fomin12, Day:2018nja}. Calculations from~\cite{fomin17, Sargsian:2019joj} suggest that high $x$ data for $Q^2 \approx 3$~GeV$^2$ may be sufficient to observe the dominance of 3N-SRCs, but the existing data do not provide a sufficiently precise measure of the $x$ dependence and give no information on the $Q^2$ dependence~\cite{fomin12}. As such, we do not yet know whether it is possible to isolate the presence of 3N-SRCs in inclusive scattering~\cite{Arrington:2022sov, Fomin:2023gdz}.

The comparison of $A=3$ mirror nuclei provides unique advantages in probing 3N-SRCs. In conventional searches for 3N-SRCs, a heavy nucleus which can contain four different 3N isospin configurations ($nnp$, $ppn$, $nnn$, $ppp$) is compared to the $npp$ SRC in $^3$He, and the 3N-SRCs in the heavy nucleus can have CM motion that is not present in $^3$He. In the $^3$H/$^3$He comparison, the 3N-SRC configurations have zero total momentum, and only a single isospin configuration is possible for each nucleus. We also expect an expanded scaling region for the $^3$H/$^3$He ratios because (a) light nuclei have smaller Fermi motion, so the MF contributions will drop off at lower $k_{min}$ values and (b) the scaling violation effects such as long-range FSIs and CM motion of the SRCs should be nearly identical in mirror nuclei. Finally, only 2N- and 3N-SRCs can contribute to the cross section at very large $x$, so any deviation from the ratio in the 2N-SRC regime should reflect the 3N-SRC contributions, even if the 2N-SRC contributions are not completely negligible. This provides sensitivity to the structure of the 3N-SRCs in $^3$H and $^3$He even if we cannot demonstrate 3N-SRC dominance.

In this work, we examine 2N-SRC scaling behaviors in $^3$H/$^2$H, and $^3$He/$^2$H and confirm the expectation of improved scaling in these few-body nuclei. We next examine the $^3$H/$^3$He ratios and demonstrate that the 2N-SRC scaling region starts at even lower $x$ and $Q^2$ values in the comparison of these mirror nuclei. We then take advantage of the greater kinematic range of the data and study the ratios at $x>$2 to examine how the $^3$H/$^3$He ratio - sensitive to the isospin structure of 3N-SRCs - changes compared to the ratio observed in the 2N-SRC region.

\begin{table}[htb]
\begin{tabular}{c c c c c}
$E_{beam}$ & $\theta_0$    & $\langle Q^2 \rangle$         &  $x$  & Ref. \\
(GeV)     &(\textdegree) & (GeV$^2$)   &   range       &  \\ \hline
2.222     & 21.778     &  0.6        &   0.6-3  & \cite{Santiesteban:2023rsh}\\
2.222     & 23.891     &  0.7        &   0.6-2  & \cite{Santiesteban:2023rsh}\\
2.222     & 30.001     &  1.0        &   0.7-2  & \cite{Santiesteban:2023rsh}\\
4.332     & 17.006     &  1.4        &   0.6-3  & \cite{Li2022}  \\
4.325     & 20.881     &  1.9        &   0.9-1.7& \cite{Li2022}  \\
5.766     & 18.000     &  2.6        &   0.5-2.1 & \cite{fomin12}  \\
\end{tabular}
\caption{Kinematics for the data included in this analysis. $E_{beam}$ is the incoming beam energy, $\theta_0$ is the central spectrometer angle, and $\langle Q^2 \rangle$ is the mean $Q^2$ value for the data at $x=1.5$. The 0.6 to 1.0 GeV$^2$ data sets are extended from the published QE peak~\cite{Santiesteban:2023rsh} to $x>1$ region. And the 1.4 GeV$^2$ data set is extended from 2N-SRC~\cite{Li2022} to 3N-SRC region. Note that the 0.7~GeV$^2$ data set has a reduced $x$ range for deuterium target($0.7<x<1.5$) and the 1.9~GeV$^2$ data set includes measurements at 17.802 degrees to extend the data down to $x=0.7$.}
\label{tab:kinematics}
\end{table}

Data used in this analysis were collected via inclusive electron scattering from $^2$H, $^3$H, and $^3$He in Hall A at Jefferson Lab (JLab) in 2018 as part of experiments E12-11-112 and E12-14-011~\cite{e1211112, e1214011, cruz-torres19, cruz-torres20, Li2022, Santiesteban:2023rsh, Arrington:2023hht}, along with $^3$He/$^2$H ratios from~\cite{fomin12}. We include cross section ratios and uncertainties over the full kinematic range of E12-11-112, much of which has not been previously published. Four identical $25$~cm long aluminum cells were used to hold 70.8, 142.2, 85.0 (84.8)\footnote{The $Q^2$=1.4~GeV$^2$ data were taken in the Fall 2018 with a different tritium cell with an areal density of 84.8~mg/cm$^2$}, and 53.2~mg/cm$^2$ of $^1$H, $^2$H, $^3$H, and $^3$He gas at room temperature~\cite{Santiesteban:2018qwi, Arrington:2023hht}. Electrons scattered from the target nuclei and were detected in the Hall A High Resolution Spectrometers. For each spectrometer angle setting, 2-4 overlapping momentum settings were used to cover an $x$ range around the QE peak and into the SRC region at $x>$1. Detailed descriptions of the experimental setup can be found in~\cite{shujie_thesis, nathaly_thesis, Alcorn:2004sb}, and the kinematics of the data presented are summarized in Table~\ref{tab:kinematics}. The cross section ratios are included in the supplemental materials~\cite{supp}. Systematical and normalization uncertainties are discussed in~\cite{Li2022}.

\begin{figure}[htb!]
    \includegraphics[width=0.68\textwidth]{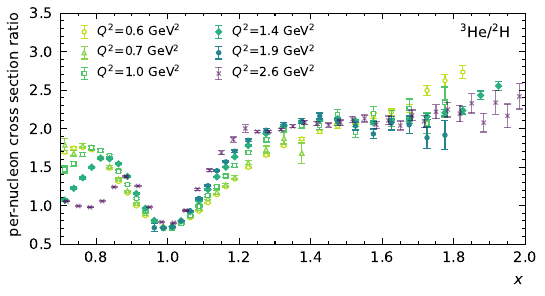}\\
    \includegraphics[width=0.68\textwidth]{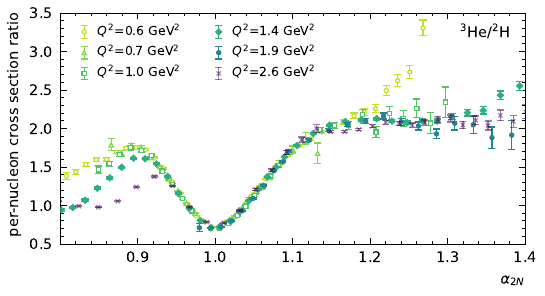}\\
    \includegraphics[width=0.68\textwidth]{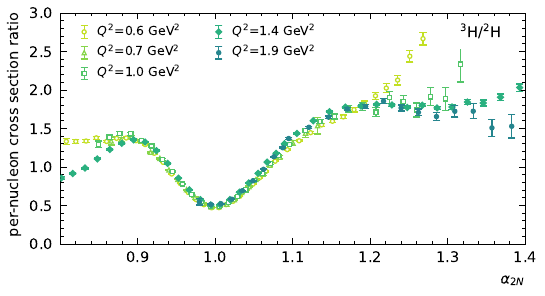}
    \caption{$^3$He/$^2$H per-nucleon cross section ratio vs $x$ (top) and $\alpha_{2N}$ (middle), along with the $^3$H/$^2$H ratios vs $\alpha_{2N}$ (bottom). Error bars indicate the statistical and systematic uncertainties summed in quadrature. The $^3$He/$^2$H ($^3$H/$^2$H) ratios have a normalization uncertainty of 1.18\% (0.78\%), except for the $Q^2=2.6$~GeV$^2$ data of $^3$He/$^2$H~\cite{fomin12} which has a 1.8\% normalization uncertainty.}
    \label{fig:3He-x-alpha}
\end{figure}

Figure~\ref{fig:3He-x-alpha} shows the $^3$He/$^2$H ratios for the full $Q^2$ range of the experiment vs $x$ (top) and $\alpha_{2N}$ (middle), and the $^3$H/$^2$H vs $\alpha_{2N}$ (bottom). The light-cone momentum $\alpha$ has a more direct connection to the struck nucleon momentum~\cite{frankfurt88, frankfurt93, Arrington:2022sov}. Reconstructing $\alpha$ in inclusive scattering requires an assumption about the nature of the final state, and for the study of 2N-SRCs we use $\alpha_{2N}$, 
\begin{equation}
    \alpha_{2N} = 2 - \frac{q_-+2m_N}{2m_N}\Big(1+\frac{\sqrt{W^2_{2N}-4m^2_N}}{W_{2N}}\Big).
\end{equation}
Here $m_N$ is the nucleon mass, $q_-=q_0-|\textbf{q}|$, $W^2_{2N}=(q+2m_N)^2=-Q^2+4q_om_N+4m^2_N$. The quantity $\alpha_{2N}$ represents the light-cone momentum fraction assuming that the momentum of the struck nucleon is balanced by a single spectator nucleon. This is appropriate for the deuteron or for scattering from a 2N-SRC with low total momentum. It is also very close to $\alpha$ calculated for the MF case, where the recoil is assumed to be taken by the whole A-1 spectator system~\cite{frankfurt88, frankfurt93, fomin17}. The benefit of using $\alpha_{2N}$ is illustrated by the fact that the clear $Q^2$ dependence in the QE peak region of the top panel of Fig.~\ref{fig:3He-x-alpha} disappears when plotted against $\alpha_{2N}$, as observed previously~\cite{fomin17, Arrington:2022sov}. As such, all further scaling studies in this work will show the ratios as a function of the estimated light-cone momentum.

In the 2N-SRC region, a plateau is seen in the $^3$H/$^2$H as well as $^3$He/$^2$H ratios at $\alpha_{2N} > 1.2$ for $Q^2$=1~GeV$^2$ and higher. This is consistent with $^4$He/$^2$H ratios of Ref.~\cite{frankfurt93}, while $^{56}$Fe/$^2$H and $^{197}$Au/$^2$H~\cite{frankfurt93} and $^{12}$C/$^3$He~\cite{egiyan03} ratios show scaling for $Q^2 \ge 1.4$~GeV$^2$. As expected, the reduced FSI contributions in $^3$H and $^3$He and more complete cancellation of FSIs in the A/$^2$H ratio lead to an expanded scaling region, with the SRC plateau observed down to $Q^2 \approx 1$~GeV$^2$. In addition, the plateau for $^3$H and $^3$He extends down to $\alpha_{2N} \approx 1.2$ ($k_{min}\approx~200$MeV), compared to $\alpha_{2N}\approx 1.3$ ($k_{min}\approx 300$~MeV) in heavier nuclei~\cite{fomin17}, due to the lower MF momenta in light nuclei.

\begin{figure}[htb]
    \includegraphics[width=0.68\textwidth]{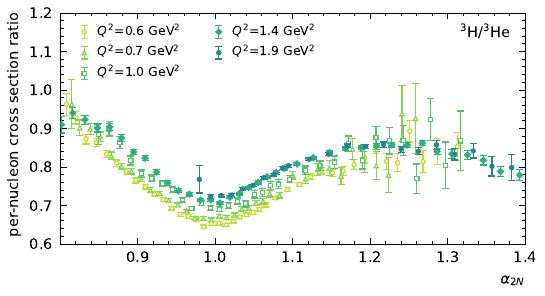}
    \caption{$^3$H/$^3$He ratios vs $\alpha_{2N}$ in the 2N-SRC region. Error bars indicate the statistical and systematic uncertainties summed in quadrature. A normalization uncertainty of 1.18\% is not shown. Note that $\alpha_{2N}$ is kinematically prohibited beyond the deuteron elastic peak $(x\approx 2)$, which is approximately $\alpha_{2N}=1.35$ and 1.43 at $Q^2$=1.4 and 1.9~GeV$^2$ respectively.}
    \label{fig:3H3He-ratios-all}
\end{figure}

The $^3$H/$^3$He ratio should show the same early onset of 2N-SRC scaling in $\alpha$, while the cancellation between FSIs in the mirror nuclei should yield a plateau at even lower $Q^2$ values. One can see this in Figure~\ref{fig:3H3He-ratios-all}, where scaling in the 2N-SRC region ($1.2< \alpha_{2N} <1.35$ in Fig.~\ref{fig:3He-x-alpha}) extends down to $Q^2=0.6$~GeV$^2$, albeit with limited statistics and limited $\alpha_{2N}$ coverage. Note that the ratio at $x \approx 1$ varies slightly with $Q^2$ because of the small difference in the $Q^2$ dependence of the proton and neutron form factors. 

Given the expanded 2N-SRC scaling region observed for $^3$H/$^3$He, it is natural to look at the ratios at $x>2$ to study the contributions from 3N-SRCs. While the $\alpha$ and $Q^2$ coverage is limited compared to some previous 3N-SRC studies~\cite{fomin12, ye18}, the expanded scaling region, combined with the simplification of having only a single, stationary 3N-SRC configuration in each nucleus, provides unique sensitivity to the 3N-SRC structure~\cite{Arrington:2022sov, Fomin:2023gdz}. 

If the three-body system is dominated by the maximally-symmetric ``triangle'' or ``star'' configurations, where all three nucleons have identical momentum, then the probability of scattering from a high-momentum proton or neutron will reflect the number of protons and neutrons in the 3N-SRC. For the kinematics of the present measurement, the ratio would approach $(\sigma_{ep} +2\sigma_{en}) / (2\sigma_{ep}+\sigma_{en}) = 0.7-0.75$, depending on $Q^2$, where $\sigma_{eN}$ are the off-shell electron-nucleon elastic scattering cross sections. The other extreme is the ``linear'' configuration, where the highest-momentum nucleon is balanced by two co-linear spectator nucleons. In this case, the cross section at the largest $\alpha$ values will be dominated by scattering from the highest-momentum nucleon. If this is the singly-occurring nucleon, then the scattering would select the proton (neutron) in $^3$H ($^3$He), yielding a $^3$H/$^3$He cross section ratio of $\sigma_{ep} / \sigma_{en} \approx 2.5$. Similarly, dominance of the doubly-occurring nucleon at large $\alpha$ would yield a ratio of $\sigma_{en} / \sigma_{ep} \approx 0.4$. If there is no isospin preference for the highest-momentum nucleon, then the scattering samples all nucleons identically, yielding a ratio of $0.7-0.75$.

While these examples assume dominance of a single configuration, they illustrate the sensitivity of the cross section ratio to the 3N-SRC isospin configuration. An analysis of $^3$He electro-disintegration suggests that the 3N-SRC is predominately formed by two consecutive 2N interactions, and that the majority of the 3N-SRCs in the nucleus are in a linear configuration~\cite{PhysRevC.71.044615}. The dominance of the linear configuration in the wavefunction is further enhanced in the cross section, as the spectator nucleons in the star configuration have larger momenta, yielding a significant excitation of the spectator system. In fact, for the kinematics of this measurement, the star configuration cannot contribute above $x \approx 1.4$ in inclusive scattering from $A=3$ nuclei~\cite{Denniston2023} due to the larger excitation energy of the spectator system~\cite{PhysRevC.71.044615}.
The spectator system in the linear configuration has a smaller excitation than in the case of 2N-SRCs for the same struck nucleon momentum, allowing this part of 3N-SRC configuration to be enhanced in the cross section relative to the 2N-SRC. Intermediate configurations between the star configuration and the minimally-symmetric linear configuration may be accessible, but will still be strongly suppressed by the increased spectator excitation. Thus, the linear (or near-linear) configurations should dominate, and we focus on the near-linear configuration when interpreting the data. Note that the predictions for isospin sensitivity are identical for linear or nearly-linear configurations, as long as the leading nucleon always has significantly higher momenta than the other two.

\begin{figure}[htb]
  \includegraphics[width=0.68\textwidth]{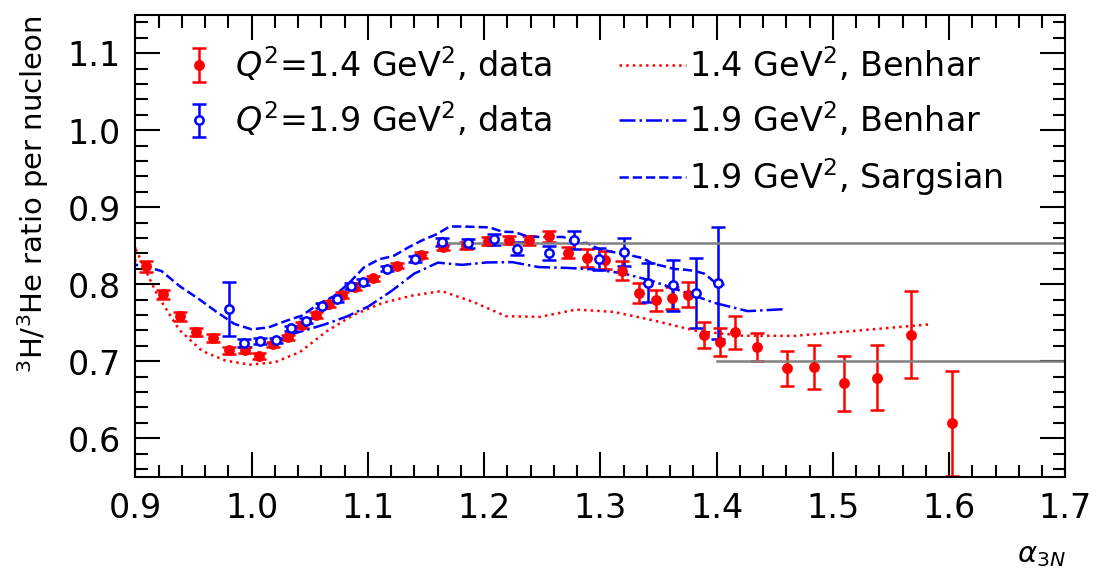}
    \caption{$^3$H/$^3$He ratios for the highest-$Q^2$ data sets. The solid lines shown for $\alpha_{3N}>$ 1.2 and 1.4 indicate the fitted value $R=0.854$ in the 2N-SRC plateau region~\cite{Li2022}, and the prediction for the 3N-SRC region, $R\approx0.7$~\cite{Day:2018nja} respectively.
    The curves are calculations from Benhar~\cite{Benhar1995, Benhar2013} and Sargsian~\cite{misak_2024, Day:2018nja}.}
    \label{fig:3H3He-ratios-highQ}
\end{figure}

Figure~\ref{fig:3H3He-ratios-highQ} shows the $^3$H/$^3$He cross section ratios vs $\alpha_{3N}$ from the two high-$Q^2$ data sets, with $x=2$ corresponding to roughly $\alpha_{3N}=1.4$. Similar to $\alpha_{2N}$,

\begin{align}
    \alpha_{3N} = 3 & - \frac{q_- + 3m_N}{2m_N} \left(1 + \frac{m_s^2 - m_N^2}{W_{3N}^2} \right.\nonumber\\ 
    & \left. + \sqrt{\left(1 - \frac{(m_s + m_N)^2}{W_{3N}^2}\right) \left(1 - \frac{(m_s - m_N)^2}{W_{3N}^2}\right)} \right)
\end{align}
approximates the light-cone momentum fraction in the three-nucleon system~\cite{Sargsian:2019joj}. Here $W_{3N} = \sqrt{\frac{Q^2 (3 - x)}{x} + 9m_N^2}$ is the invariant mass of the 3N system, and $m_s= \sqrt{4 (m_N^2 + 4k^2)}$ is the mass of the 2N spectator system where the relative momentum between two spectator nucleons $k=0$ in the linear configuration. $\alpha_{3N}$ is nearly identical to $\alpha_{2N}$ for the mean-field and 2N-SRC regions, while being more appropriate for examining 3N-SRC configurations. The dotted and dashed curves represent spectral function-based calculations of inclusive cross section ratios from~\cite{Benhar1993, Benhar2013} and~\cite{misak_2024} (based on the formalism of Ref.~\cite{Day:2018nja}. The two $Q^2=1.9$~GeV$^2$ calculations are in good qualitative agreement with the data near the QE peak and the 2N-SRC region, although they differ by $\sim$5\%. While the 1.9~GeV$^2$ calculations are in pretty good agreement with the data, the 1.4~GeV$^2$ result~\cite{Benhar2013} shows a significant difference between the ratios at 1.4 and 1.9~GeV$^2$ that is not observed in the data. 

The horizontal lines in Fig.~\ref{fig:3H3He-ratios-highQ} indicate the observed cross section ratio in the 2N-SRC regime~\cite{Li2022}, and the predicted 3N-SRC plateau value of
\begin{equation}
    a_3(^3H)\sim \left[\frac{a_2(^3H)}{a_2(^3He)}\right]^2 = \left[\frac{\sigma(^3H)}{\sigma(^3He)}\right]^2 \approx 0.7 \label{eq:a3}
\end{equation}
assuming dominance of the linear configuration from two consecutive hard two-nucleon interactions~\cite{Sargsian:2019joj, Day:2018nja}. Ref.~\cite{Day:2018nja} predicted the onset of 3N-SRC dominance somewhere above $\alpha_{3N} = 1.6$ ($k_{min} \approx 600$~MeV) based on the requirement that both of the spectators be above $k_{min} \approx 300$~MeV to exclude MF contributions. For A=3 nuclei, 2N-SRC dominance is observed above $\alpha=1.2$ ($k_{min} \approx 200$~MeV), so the same criteria imply possible 3N-SRCs dominate above $\alpha \approx 1.4$ ($k_{min}=400$~MeV, corresponding to a linear configuration with two 200~MeV spectators). In addition, calculations~\cite{Day:2018nja, Sargsian:2019joj} show that two-body breakup, FSIs, and 2N-SRCs are suppressed in the cross section ratios in that region. That means the present data may be sufficient to isolate 3N-SRCs in A=3.
For a more detailed treatment and analysis of the various contributions, see Ref.~\cite{Day:2018nja} 

The ratio at $\alpha_{3N} > 1.4$ is consistent with a plateau near 0.7, as predicted by Eq.~\ref{eq:a3}. If 3N-SRCs dominate, this suggests an isospin-symmetric configuration. Even if these kinematics do not fully isolate 3N-SRCs, it is clear that the 3N-SRC contribution decreases the ratio below the value in the 2N-SRC region, indicating that the contribution from the 3N-SRCs is at or below the observed value of 0.7, ruling out configurations where the doubly-occurring nucleon dominates at the largest momenta. Additional measurements, extending the $\alpha_{3N}$ range and providing measurements at more than one $Q^2$ value, are needed to have a complete test of the prediction of 3N-SRC dominance.

In conclusion, we present new measurements of inclusive cross section ratios in kinematics corresponding to scattering from high-momentum nucleons in SRCs over a range of $Q^2$. We examined the behavior at lower $Q^2$ values, where the $x$-scaling predicted by the SRC model has been observed to break down in previous A/$^2$H ratio measurements, and found that the ratios of $^3$He and $^3$H to $^2$H have an expanded scaling region in both $x$ ($\alpha$) and $Q^2$ compared to ratios on heavier nuclei. This earlier dominance of the SRCs results from the reduced MF momenta in these nuclei combined with smaller scaling violations associated with FSI or CM motion of the SRCs in the ratios. Specifically, the FSIs and CM motion contributions are both small in absolute sizes, and largely cancel in comparisons of these light nuclei. That leads to the observed 2N-SRC scaling down to $Q^2$=0.6~GeV$^2$ in $^3$H/$^3$He ratios.

This also represents the first extraction of the $^3$H/$^3$He ratio beyond the 2N-SRC region, allowing us to examine the behavior of the cross section ratio as we move into kinematics where 3N-SRC contributions are expected. While this measurement is limited to $\alpha \leq 1.6$, too low to isolate 3N-SRC in heavy nuclei according to recent predictions~\cite{Sargsian:2019joj, Day:2018nja}, it may be sufficient to isolate 3N-SRCs in A=3 nuclei~\cite{misak_2024}. The criteria of Ref.~\cite{Sargsian:2019joj}, evaluated using the smaller Fermi momenta for A=3 nuclei, suggest that 3N-SRCs may begin to dominate above $\alpha=1.4$. Our data are consistent with a constant ratio for $\alpha \gtorder 1.45$, and the value of this ratio is consistent with isospin-independent 3N-SRC configurations. Even if 3N-SRC is not dominant in that region, the decrease of the ratio from the 2N-SRC plateau of 0.84 rules out dominance of the singly-occurring nucleon at the largest momenta in 3N-SRCs, which would yield a significant increase in the ratio. A more definitive test of scaling and a clearer picture of the isospin structure of 3N-SRCs will require $^3$H/$^3$He inclusive ratio with high precision at the largest $\alpha_{3N}$ and measurements at multiple $Q^2$ values.


We acknowledge useful discussions with M. Sargsian, O. Benhar and are grateful to the Jefferson Lab target group and technical staff for the design and construction of the Tritium target and their support running this experiment. This work was supported in part by the Department of Energy's Office of Science, Office of Nuclear Physics, under contracts DE-AC02-05CH11231, DE-FG02-88ER40410, DE-SC0014168, and DE-FG02-96ER40950, the National Science Foundation including grant NSF PHY-1714809, and DOE contract DE-AC05-06OR23177 under which JSA, LLC operates JLab. Z.H.Y. acknowledges the support from the National Science Foundation of China under contract 12275148. A.S. acknowledges the support from the Science Committee of the Republic of Armenia under grant 21AG-1C085.


\bibliographystyle{elsarticle-num}
\bibliography{tritium_SRC}

\begin{thebibliography}{10}
\expandafter\ifx\csname url\endcsname\relax
  \def\url#1{\texttt{#1}}\fi
\expandafter\ifx\csname urlprefix\endcsname\relax\def\urlprefix{URL }\fi
\expandafter\ifx\csname href\endcsname\relax
  \def\href#1#2{#2} \def\path#1{#1}\fi

\bibitem{frankfurt88}
L.~Frankfurt, M.~Strikman, {Hard Nuclear Processes and Microscopic Nuclear Structure}, Phys. Rept. 160 (1988) 235.
\newblock \href {https://doi.org/10.1016/0370-1573(88)90179-2} {\path{doi:10.1016/0370-1573(88)90179-2}}.

\bibitem{Arrington:2022sov}
J.~Arrington, N.~Fomin, A.~Schmidt, {Progress in Understanding Short-Range Structure in Nuclei: An Experimental Perspective}, Annual Review of Nuclear and Particle Science 72 (2022) 307.

\bibitem{sargsian03}
M.~M. Sargsian, et~al., {Hadrons in the nuclear medium}, J. Phys. G29 (2003) R1.
\newblock \href {https://doi.org/10.1088/0954-3899/29/3/201} {\path{doi:10.1088/0954-3899/29/3/201}}.

\bibitem{arrington12a-SRCreview}
J.~Arrington, D.~Higinbotham, G.~Rosner, M.~Sargsian, Hard probes of short-range nucleon–nucleon correlations, Prog. Part. Nucl. Phys. 67 (2012) 898.
\newblock \href {https://doi.org/10.1016/j.ppnp.2012.04.002} {\path{doi:10.1016/j.ppnp.2012.04.002}}.

\bibitem{frankfurt93}
L.~L. Frankfurt, M.~I. Strikman, D.~B. Day, M.~Sargsyan, Evidence for short-range correlations from high ${\mathit{q}}^{2}$ ( \textit{e} , \textit{e} ') reactions, Phys. Rev. C 48 (1993) 2451.
\newblock \href {https://doi.org/10.1103/PhysRevC.48.2451} {\path{doi:10.1103/PhysRevC.48.2451}}.

\bibitem{fomin12}
N.~Fomin, et~al., {New measurements of high-momentum nucleons and short-range structures in nuclei}, Phys. Rev. Lett. 108 (2012) 092502.
\newblock \href {https://doi.org/10.1103/PhysRevLett.108.092502} {\path{doi:10.1103/PhysRevLett.108.092502}}.

\bibitem{schmookler19}
B.~Schmookler, et~al., {Modified structure of protons and neutrons in correlated pairs}, Nature 566 (2019) 354.
\newblock \href {https://doi.org/10.1038/s41586-019-0925-9} {\path{doi:10.1038/s41586-019-0925-9}}.

\bibitem{Li2022}
S.~Li, R.~Cruz-Torres, N.~Santiesteban, Z.~H. Ye, et~al., {Revealing the short-range structure of the mirror nuclei 3H and 3He}, Nature 609 (2022) 41.
\newblock \href {https://doi.org/10.1038/s41586-022-05007-2} {\path{doi:10.1038/s41586-022-05007-2}}.

\bibitem{egiyan03}
K.~S. Egiyan, et~al., {Observation of nuclear scaling in the A(e, e-prime) reaction at x(B) greater than 1}, Phys. Rev. C 68 (2003) 014313.
\newblock \href {https://doi.org/10.1103/PhysRevC.68.014313} {\path{doi:10.1103/PhysRevC.68.014313}}.

\bibitem{fomin17}
N.~Fomin, D.~Higinbotham, M.~Sargsian, P.~Solvignon, {New Results on Short-Range Correlations in Nuclei}, Ann. Rev. Nucl. Part. Sci. 67 (2017) 129--159.
\newblock \href {https://doi.org/10.1146/annurev-nucl-102115-044939} {\path{doi:10.1146/annurev-nucl-102115-044939}}.

\bibitem{Atti1994}
C.~degli Atti, S.~Simula, {Nucleon-nucleon correlations and final state interaction in inclusive quasi-elastic electron scattering off nuclei at x > 1}, Physics Letters B 325~(3-4) (1994) 276--282.
\newblock \href {http://arxiv.org/abs/9403001} {\path{arXiv:9403001}}, \href {https://doi.org/10.1016/0370-2693(94)90010-8} {\path{doi:10.1016/0370-2693(94)90010-8}}.

\bibitem{Benhar1995}
O.~Benhar, A.~Fabrocini, S.~Fantoni, I.~Sick, {Inclusive cross section ratios at x >1}, Physics Letters B 343 (1995) 47.
\newblock \href {https://doi.org/10.1016/0370-2693(94)01468-R} {\path{doi:10.1016/0370-2693(94)01468-R}}.

\bibitem{Frankfurt2008}
L.~Frankfurt, M.~Sargsian, M.~Strikman, Recent observation of short-range nucleon correlations in nuclei and their implication for the structure of nuclei and neutron stars, International Journal of Modern Physics A 23 (2008) 2991.

\bibitem{Shneor:2007tu}
R.~Shneor, et~al., {Investigation of proton-proton short-range correlations via the C-12(e, e-prime pp) reaction}, Phys. Rev. Lett. 99 (2007) 072501.
\newblock \href {https://doi.org/10.1103/PhysRevLett.99.072501} {\path{doi:10.1103/PhysRevLett.99.072501}}.

\bibitem{Subedi:2008zz}
R.~Subedi, et~al., {Probing Cold Dense Nuclear Matter}, Science 320 (2008) 1476--1478.
\newblock \href {https://doi.org/10.1126/science.1156675} {\path{doi:10.1126/science.1156675}}.

\bibitem{Korover:2014dma}
I.~Korover, et~al., {Probing the Repulsive Core of the Nucleon-Nucleon Interaction via the $^4He(e,e'pN)$ Triple-Coincidence Reaction}, Phys. Rev. Lett. 113 (2014) 022501.
\newblock \href {https://doi.org/10.1103/PhysRevLett.113.022501} {\path{doi:10.1103/PhysRevLett.113.022501}}.

\bibitem{CLAS:2018xvc}
M.~Duer, et~al., {Direct Observation of Proton-Neutron Short-Range Correlation Dominance in Heavy Nuclei}, Phys. Rev. Lett. 122 (2019) 172502.
\newblock \href {https://doi.org/10.1103/PhysRevLett.122.172502} {\path{doi:10.1103/PhysRevLett.122.172502}}.

\bibitem{Nguyen:2020mgo}
D.~Nguyen, et~al., {Novel observation of isospin structure of short-range correlations in calcium isotopes}, Phys. Rev. C 102 (2020) 064004.
\newblock \href {https://doi.org/10.1103/PhysRevC.102.064004} {\path{doi:10.1103/PhysRevC.102.064004}}.

\bibitem{schiavilla07}
R.~Schiavilla, R.~B. Wiringa, S.~C. Pieper, J.~Carlson, {Tensor Forces and the Ground-State Structure of Nuclei}, Phys. Rev. Lett. 98 (2007) 132501.
\newblock \href {https://doi.org/10.1103/PhysRevLett.98.132501} {\path{doi:10.1103/PhysRevLett.98.132501}}.

\bibitem{alvioli08}
M.~Alvioli, C.~Ciofi~degli Atti, H.~Morita, {Proton-neutron and proton-proton correlations in medium-weight nuclei and the role of the tensor force}, Phys. Rev. Lett. 100 (2008) 162503.
\newblock \href {https://doi.org/10.1103/PhysRevLett.100.162503} {\path{doi:10.1103/PhysRevLett.100.162503}}.

\bibitem{wiringa14}
{Wiringa, R.B. and Schiavilla, R. and Pieper, Steven C. and Carlson, J.}, {Nucleon and nucleon-pair momentum distributions in $A \le 12$ nuclei}, Phys. Rev. C 89 (2014) 024305.
\newblock \href {https://doi.org/10.1103/PhysRevC.89.024305} {\path{doi:10.1103/PhysRevC.89.024305}}.

\bibitem{Terashima:2018bwq}
S.~Terashima, et~al., {Dominance of tensor correlations in high-momentum nucleon pairs studied by (p,pd) reaction}, Phys.\ Rev.\ Lett. 121 (2018) 242501.
\newblock \href {http://arxiv.org/abs/1811.02118} {\path{arXiv:1811.02118}}, \href {https://doi.org/10.1103/PhysRevLett.121.242501} {\path{doi:10.1103/PhysRevLett.121.242501}}.

\bibitem{arrington12b-EMCSRC}
J.~Arrington, A.~Daniel, D.~Day, N.~Fomin, D.~Gaskell, P.~Solvignon, {A detailed study of the nuclear dependence of the EMC effect and short-range correlations}, Phys. Rev. C 86 (2012) 065204.
\newblock \href {https://doi.org/10.1103/PhysRevC.86.065204} {\path{doi:10.1103/PhysRevC.86.065204}}.

\bibitem{Weiss:2020bkp}
R.~Weiss, A.~W. Denniston, J.~R. Pybus, O.~Hen, E.~Piasetzky, A.~Schmidt, L.~B. Weinstein, N.~Barnea, {Extracting the number of short-range correlated nucleon pairs from inclusive electron scattering data}, Phys. Rev. C 103 (2021) L031301.
\newblock \href {https://doi.org/10.1103/PhysRevC.103.L031301} {\path{doi:10.1103/PhysRevC.103.L031301}}.

\bibitem{Ye:2018jth}
Z.~Ye, J.~Arrington, {Inclusive Studies of Short-Range Correlations: Overview and New Results}, in: {13th Conference on the Intersections of Particle and Nuclear Physics}, 2018.
\newblock \href {http://arxiv.org/abs/1810.03667} {\path{arXiv:1810.03667}}.

\bibitem{Fomin:2023gdz}
N.~Fomin, J.~Arrington, S.~Li, {Searching for three-nucleon short-range correlations}, The European Physical Journal A 59 (2023) 205.
\newblock \href {https://doi.org/10.1140/epja/s10050-023-01112-6} {\path{doi:10.1140/epja/s10050-023-01112-6}}.

\bibitem{Day:2018nja}
D.~B. Day, L.~L. Frankfurt, M.~M. Sargsian, M.~I. Strikman, {Toward observation of three-nucleon short-range correlations in high-Q2~A(e,e')X reactions}, Phys. Rev. C 107 (2023) 014319.
\newblock \href {https://doi.org/10.1103/PhysRevC.107.014319} {\path{doi:10.1103/PhysRevC.107.014319}}.

\bibitem{ye18}
Z.~Ye, et~al., {Search for three-nucleon short-range correlations in light nuclei}, Phys. Rev. C 97 (2018) 065204.
\newblock \href {https://doi.org/10.1103/PhysRevC.97.065204} {\path{doi:10.1103/PhysRevC.97.065204}}.

\bibitem{Sargsian:2019joj}
M.~M. Sargsian, D.~B. Day, L.~L. Frankfurt, M.~I. Strikman, {Searching for three-nucleon short-range correlations}, Phys. Rev. C 100 (2019) 044320.
\newblock \href {https://doi.org/10.1103/PhysRevC.100.044320} {\path{doi:10.1103/PhysRevC.100.044320}}.

\bibitem{Santiesteban:2023rsh}
S.~N. Santiesteban, et~al., {Novel Measurement of the Neutron Magnetic Form Factor from A=3 Mirror Nuclei}, Phys. Rev. Lett. 132 (2024) 162501.
\newblock \href {https://doi.org/10.1103/PhysRevLett.132.162501} {\path{doi:10.1103/PhysRevLett.132.162501}}.

\bibitem{e1211112}
J.~Arrington, D.~Day, D.~W. Higinbotham, P.~Solvignon, {Precision measurement of the isospin dependence in the 2N and 3N short range correlation region}, Jefferson Lab Experiment Proposal E12-11-112 (2011).

\bibitem{e1214011}
L.~Weinstein, W.~Boeglin, O.~Chen, F.~Hauenstein, {Proton and Neutron Momentum Distributions in A = 3 Asymmetric Nuclei}, Jefferson Lab Experiment Proposal E12-14-011 (2014).

\bibitem{cruz-torres19}
R.~Cruz-Torres, et~al., {Comparing proton momentum distributions in $A=2$ and 3 nuclei via $^2$H $^3$H and $^3$He $(e, e'p)$ measurements}, Phys. Lett. B 797 (2019) 134890.
\newblock \href {https://doi.org/10.1016/j.physletb.2019.134890} {\path{doi:10.1016/j.physletb.2019.134890}}.

\bibitem{cruz-torres20}
R.~Cruz-Torres, et~al., {Probing Few-Body Nuclear Dynamics via $^3$H and $^3$He ($e,e'p$)pn Cross-Section Measurements}, Phys. Rev. Lett. 124 (2020) 212501.
\newblock \href {https://doi.org/10.1103/PhysRevLett.124.212501} {\path{doi:10.1103/PhysRevLett.124.212501}}.

\bibitem{Arrington:2023hht}
J.~Arrington, R.~Cruz-Torres, T.~J. Hague, L.~Kurbany, S.~Li, D.~Meekins, N.~Santiesteban, {The Jefferson Lab tritium program of nucleon and nuclear structure measurements}, Eur. Phys. J. A 59 (2023) 188.
\newblock \href {https://doi.org/10.1140/epja/s10050-023-01085-6} {\path{doi:10.1140/epja/s10050-023-01085-6}}.

\bibitem{Santiesteban:2018qwi}
S.~N. Santiesteban, et~al., {Density Changes in Low Pressure Gas Targets for Electron Scattering Experiments}, Nucl. Instrum. Meth. A 940 (2019) 351.
\newblock \href {https://doi.org/10.1016/j.nima.2019.06.025} {\path{doi:10.1016/j.nima.2019.06.025}}.

\bibitem{shujie_thesis}
S.~Li, {Ph.D Thesis, University of New Hampshire} (2020).

\bibitem{nathaly_thesis}
S.~Santiesteban, {Ph.D Thesis, University of New Hampshire} (2020).

\bibitem{Alcorn:2004sb}
J.~Alcorn, et~al., {Basic Instrumentation for Hall A at Jefferson Lab}, Nucl. Instrum. Meth. A 522 (2004) 294.
\newblock \href {https://doi.org/10.1016/j.nima.2003.11.415} {\path{doi:10.1016/j.nima.2003.11.415}}.

\bibitem{supp}
See Supplemental Material at [URL will be inserted by publisher] for detailed expressions and tables of cross section results. (2025).

\bibitem{PhysRevC.71.044615}
M.~M. Sargsian, T.~V. Abrahamyan, M.~I. Strikman, L.~L. Frankfurt, Exclusive electrodisintegration of $^{3}\mathrm{He}$ at high ${Q}^{2}$. ii. decay function formalism, Phys. Rev. C 71 (2005) 044615.
\newblock \href {https://doi.org/10.1103/PhysRevC.71.044615} {\path{doi:10.1103/PhysRevC.71.044615}}.

\bibitem{Denniston2023}
A.~Denniston, Search for 3n src at clas12, \url{https://indico.mit.edu/event/656/contributions/1775/}, presendted at the 4th International Workshop on Quantitative Challenges in Short-Range Correlations and the EMC Effect Research held at CEA Paris-Saclay, February 2023 (2023).

\bibitem{Benhar2013}
O.~Benhar, Final-state interactions in the nuclear response at large momentum transfer, Phys. Rev. C 87 (2013) 024606.
\newblock \href {https://doi.org/10.1103/PhysRevC.87.024606} {\path{doi:10.1103/PhysRevC.87.024606}}.

\bibitem{misak_2024}
M.~M. Sargsian, private communication (Sep. 2024).

\bibitem{Benhar1993}
O.~Benhar, V.~R. Pandharipande, Scattering of gev electrons by light nuclei, Phys. Rev. C 47 (1993) 2218.
\newblock \href {https://doi.org/10.1103/PhysRevC.47.2218} {\path{doi:10.1103/PhysRevC.47.2218}}.

\end{thebibliography}

\end{document}